\documentclass[pdflatex,sn-mathphys-num]{sn-jnl}

\usepackage{graphicx}
\usepackage{booktabs}
\usepackage{multirow}
\usepackage{tabularx}
\usepackage{makecell}
\usepackage[caption=false]{subfig}

\usepackage{amsmath,amssymb,amsfonts}
\usepackage{amsthm}
\usepackage{mathrsfs}
\usepackage{bm}
\usepackage{bbm}
\usepackage{dsfont}
\usepackage{latexsym}
\usepackage{leftidx}

\usepackage{tikz}
\usepackage{tikz-cd}
\usetikzlibrary{cd,arrows.meta,shapes.geometric,arrows}
\tikzstyle{arrow}     = [thick,->,>=stealth]
\tikzstyle{startstop} = [rectangle, rounded corners, minimum width=3cm, minimum height=1cm,
                         text centered, draw=black, fill=white!30]
\tikzstyle{process}   = [rectangle, minimum width=3cm, minimum height=1cm,
                         text centered, draw=black, fill=orange!30]
\tikzstyle{decision}  = [diamond, minimum width=3cm, minimum height=1cm,
                         text centered, draw=black, fill=green!30]

\usepackage{algorithm}
\usepackage{algorithmicx}
\usepackage{algpseudocode}
\usepackage{listings}

\usepackage[utf8]{inputenc} % pdflatex 时保留
\usepackage{textcomp}
\usepackage{xcolor}
\usepackage{url}
\usepackage{microtype}
\usepackage{comment}
\usepackage{manyfoot}
\usepackage[title]{appendix}
\usepackage{titlesec}

\usepackage[normalem]{ulem}

\usepackage[intoc]{nomencl}
\makenomenclature

\usepackage[symbols,nogroupskip,sort=none]{glossaries-extra}

\usepackage{xr}
\usepackage{xcite}

\usepackage{epstopdf}
\usepackage{psfrag}
\usepackage{pst-all}
\usepackage[off]{auto-pst-pdf}

\usepackage{feynmp}
\usepackage{slashed}

\definecolor{darkblue}{rgb}{0.,0.,0.4}
\definecolor{darkred}{rgb}{0.5,0.,0.}
\definecolor{BlueViolet}{RGB}{138,43,226}
\definecolor{SkyBlue}{RGB}{30,144,255}
\definecolor{DarkGreen}{RGB}{0,100,0}

\makeatletter
\newcommand*{\addFileDependency}[1]{%
  \typeout{(#1)}%
  \@addtofilelist{#1}%
  \IfFileExists{#1}{}{\typeout{No file #1.}}%
}
\makeatother

\newcommand*{\myexternaldocument}[1]{%
  \externaldocument{#1}%
  \addFileDependency{#1.tex}%
  \addFileDependency{#1.aux}%
}

\myexternaldocument{main_05_12}

\theoremstyle{thmstyleone}%
\newtheorem{theorem}{Theorem}
\newtheorem{proposition}[theorem]{Proposition}
\newtheorem{corollary}{Corollary}

\theoremstyle{thmstyletwo}%
\newtheorem{example}{Example}
\newtheorem{remark}{Remark}
\newtheorem{question}{Question}

\theoremstyle{thmstylethree}%
\newtheorem{definition}{Definition}

\usepackage{etoolbox}
\AtBeginEnvironment{theorem}{\itshape\normalsize}
\AtBeginEnvironment{proposition}{\itshape\normalsize}
\AtBeginEnvironment{lemma}{\itshape\normalsize}
\AtBeginEnvironment{corollary}{\itshape\normalsize}
\AtBeginEnvironment{conjecture}{\itshape\normalsize}
\AtBeginEnvironment{claim}{\itshape\normalsize}

\AtBeginEnvironment{definition}{\normalfont\normalsize}
\AtBeginEnvironment{example}{\normalfont\normalsize}
\AtBeginEnvironment{remark}{\normalfont\normalsize}
\AtBeginEnvironment{question}{\normalfont\normalsize}
\numberwithin{equation}{section}
\numberwithin{corollary}{section}
\numberwithin{theorem}{section}
\numberwithin{lemma}{section}
\numberwithin{definition}{section}
\numberwithin{example}{section}
\numberwithin{remark}{section}
\numberwithin{conjecture}{section}
\numberwithin{question}{section}
\numberwithin{claim}{section}

\newcommand{\beq}{\begin{eqnarray}}
\newcommand{\eeq}{\end{eqnarray}}

\newcommand{\bsp}{\begin{aligned}}
\newcommand{\esp}{\end{aligned}}

\newcommand{\ie}{{i.e., }}

\newcommand{\diff}{\mathrm{diff}}
\newcommand{\rH}{\mathrm{H}}
\newcommand{\R}{\mathbb{R}}
\newcommand{\Aut}{\mathrm{Aut}}

\newcommand{\dist}{\mathrm{dist}}

\newcommand{\im}{\mathrm{im}}

\newcommand{\ovl}{\overline}
\newcommand{\cB}{\mathscr{B}}

\newcommand{\rd}{\mathrm{d}}
\newcommand{\rmi}{\mathrm{i}}

\newcommand{\Q}{\mathbb{Q}}

\newcommand{\z}{\mathbb{Z}}

\newcommand{\A}{\mathscr{A}}

\newcommand{\G}{\mathscr{G}}
\newcommand{\cH}{\mathcal{H}}
\newcommand{\cU}{\mathscr{U}}
\newcommand{\Ad}{\mathrm{Ad}}

\newcommand{\bbC}{\mathbb{C}}

\newcommand{\rHom}{\mathrm{Hom}}
\newcommand{\QCA}{\mathrm{QCA}}

\newcommand{\id}{\mathrm{id}}

\newcommand{\cSU}{\mathscr{SU}}

\def\U{\mathrm{U}(1)}

\begin{document}

\title{Anomalies in quantum spin systems and Nielsen–Ninomiya type Theorems}

\author*[1,2]{\fnm{Ruizhi} \sur{Liu}}

\affil*[1]{\orgdiv{Department of mathematics}, \orgname{Dalhousie University}, \orgaddress{\street{6297 Castine Way}, \city{Halifax}, \postcode{B3H 4R2}, \state{Nova Scotia}, \country{Canada}}}

\affil[2]{ \orgname{Perimeter Institute for Theoretical Physics}, \orgaddress{\street{31 Caroline St N}, \city{Waterloo}, \postcode{N2L 2Y5}, \state{Ontario}, \country{Canada}}}

\abstract{We provide an algebraic perspective of Nielsen–Ninomiya–type no-go theorems arising from group-cohomological anomalies, revisiting in particular the version proved by Kapustin and Sopenko in Ref.~\cite{kapustin2024anomalous}. Departing from their analytic proof, our approach emphasizes the algebraic structure of symmetry actions and the local computability of anomaly indices. We demonstrate that this no-go theorem is due to a fundamental algebraic incompatibility between anomaly data and the dimension of local Hilbert spaces. Specifically, when an anomaly index is locally computable via quasi-local unitary operators, a suitable gauge fixing trivializes their (generalized) determinants, imposing unexpected and nontrivial constraints on lattice regularizations.}


\maketitle
\tableofcontents

\section{Introduction}
The relation between continuum quantum field theories (QFTs) and lattice models has long played a central role in both high-energy physics and condensed matter physics. From the perspective of QFT, a fundamental question is whether a given continuum theory admits a lattice regularization.

It is well known that such a regularization may not exist under some natural assumptions. A seminal example is provided by the work of Nielsen and Ninomiya \cite{NIELSEN1981219,NIELSEN198120,NIELSEN1981173}, later by Friedan \cite{Friedan1982}. More precisely, the Nielsen--Ninomiya theorem forbids a lattice regularization of \emph{free fermions} with exact lattice $\U_{V}$ charge symmetry and $\U_A$ axial (a.k.a chiral) symmetry, assuming lattice translation invariance and suitable continuity conditions on the energy spectrum. See \cite{L_scher_2002,kaplan2024chiralgaugetheoryboundary,degrand2025ginspargwilsonrelationoverlapfermions,Chatterjee_2025, gioia2025exactchiralsymmetries31d, thorngren2026chirallatticegaugetheories} for numerous proposals attempting to evade the Nielsen--Ninomiya no-go theorem.

In this work, we do \textbf{not} attempt to resolve the long-standing problem of realizing chiral fermions on the lattice. Rather, our aim is to provide an algebraic and conceptual understanding Nielsen–Ninomiya type theorems, at least in the cases arising from group cohomological anomalies.

Although the original Nielsen--Ninomiya theorem was formulated under rather restrictive assumptions—most notably the absence of interactions and the presence of lattice translation invariance—it was already recognized in Refs.~\cite{NIELSEN198120,NIELSEN1981173,NIELSEN1981219} that its underlying obstruction has a deeper origin in chiral anomalies. It is now well understood (at least in lattice systems) that anomalies are determined solely by the symmetry action and are insensitive to microscopic details of the interactions \cite{Else_2014,Cheng2022,kapustin2024anomalous,rubio2024classifyingsymmetricsymmetrybrokenspin,Liu2024LRLSM,Long_2025,kawagoe2025anomalydiagnosissymmetryrestriction,kapustin2025highersymmetriesanomaliesquantum,kapustin2025highersymmetriesanomaliescrossed,Tu2025,liu2025twistedlocalitypreservingautomorphismsanomaly,czajka2025anomalieslatticehomotopyquantum,Kobayashi_2026,Feng_2026}. 

From this viewpoint, the Nielsen--Ninomiya obstruction is largely independent of the detailed form of the Hamiltonian and does not fundamentally rely on translation symmetry.

In Ref.~\cite{jones20191dlatticemodelsboundary, Ellison_2019, czajka2025anomalieslatticehomotopyquantum}, it is argued that certain anomalies beyond group cohomology in fermionic systems cannot be realized in a lattice fermion system with tensor-factorized Hilbert spaces. 
Recently in Ref.~\cite{kapustin2024anomalous}, Kapustin and Sopenko proved an analogue of the Nielsen--Ninomiya theorem for quantum spin chains. Let $G$ be a compact, connected Lie group. They showed that $(1+1)d$ quantum field theories with anomalous $G$ symmetry cannot be regularized by spin chains with \emph{exact} $G$ symmetry, even if the symmetry action is allowed to possess decaying tails and the anomaly of $G$ symmetry lies in the group cohomology $\rH^{3}(G;\U)$, see Appendix.~\ref{sec:group_cohomology} for a brief review on group cohomology\footnote{For Lie group $G$, here by $\rH^{*}(G;\U)$ we mean the \textit{differentiable group cohomology}. See Appendix.~\ref{sec:differentiable_group_cohomology}.}. Note that  we adopt the convention that, throughout this work, a spin system is a tensor-factorized system with uniformly bounded local Hilbert space dimension.

As a corollary of the Kapustin-Sopenko's no-go theorem, the boundary theory of a bosonic integer quantum Hall state, corresponding to $G=\U$, cannot be realized in a spin chain with the correct $\U$ anomaly.

The proof in \cite{kapustin2024anomalous} relies on hard analysis, which may limit its accessibility to the broader physics community. Furthermore, it is not evident whether the underlying strategy extends to higher-dimensional spin systems or to higher-form symmetries.

In this work, we present a simpler and purely algebraic proof of this result, and develop an algebraic perspective on Nielsen--Ninomiya type theorems. This algebraic approach is conceptual, and we expect it to extend naturally to arbitrary spatial dimensions as well as to higher-form symmetries in spin systems.

We achieve this by extending the recent classification of anomalies in quantum spin systems proposed in Ref.~\cite{Tu2025} to incorporate Lie group symmetries and symmetry actions with tails in $(1+1)d$.
Conventionally, the ’t Hooft anomalies of a global symmetry $G$ in $(d+1)$-dimensional quantum spin models are believed to be classified by the group cohomology $\rH^{d+2}(G;\U)$, at least when $d$ is small. 
However, based on the homotopy theory of quantum cellular automata (QCAs)\footnote{These are automorphisms on the spin system which preserve the strict notion of locality, i.e. it maps local operators to local operators with finite spread of supports, see Def.~\ref{def:QCA} for more precise definition.}, a refined classification in spin systems was proposed in Ref.~\cite{Tu2025}. 
In this proposal, the anomaly is classified by $\rH^{d+2}(G;\Q/\z)$, i.e. the torsion subgroup of $\rH^{d+2}(G;\U)$.
In particular, since $\rH^{3}(\U;\Q/\z)\simeq 0$, the generator of $\rH^{3}(\U;\U)\simeq \z$ cannot be realized in spin chains as exact $\U$ symmetry implemented by QCAs.

This homotopical classification was further justified and developed in Ref.~\cite{czajka2025anomalieslatticehomotopyquantum}, where a detailed mathematical construction of the $\Omega$-spectrum of QCAs was presented.
While mathematically precise, the use of homotopy theory may be less accessible to a broad physics audience.

More importantly, Refs.~\cite{Tu2025,czajka2025anomalieslatticehomotopyquantum} assume that symmetry actions are implemented by QCAs, \ie strictly locality-preserving automorphisms, thereby excluding symmetry actions with decaying tails.
As pointed out in Ref.~\cite{PIRSA_Zhang}, such tails are essential for continuous symmetries; in particular, Chern insulators cannot be disentangled without allowing such tails.
This naturally raises the question of whether the $\rH^{d+2}(G;\Q/\z)$ classification of anomalies remains valid if symmetry actions with tails are taken into account.

In this work, we show that this classification is indeed robust against tails in $(1+1)d$, and we explain why the same argument is expected to extend straightforwardly to arbitrary spatial dimensions and for higher form symmetries.

More precisely, an informal version of our main result is
\begin{theorem}\label{thm:main}
    Given a spin chain with local Hilbert space dimension $n$, then the 't Hooft anomaly of $G$-symmetry action on this spin chain is classified by $\rH^{3}(G;\z[n^{-1}]/\z)$, where
    \beq
    \z[n^{-1}]=\text{the polynomial ring in $n^{-1}$ over $\z$}
    \eeq
    is the (algebraic) localization\footnote{We will write $P_{n}^{-1}$ for the localization functor for $P_{n}$, e.g. $\z[n^{-1}]:=P_{n}^{-1}\z$, see e.g. Sec.~4.7 of Ref.~\cite{rotman2008introduction}.} of $\z$ at the multiplicative subset $P_{n}:=\{n^{k}\}_{k\in\mathbb{N}^*}$. 
\end{theorem}
See Theorem \ref{thm: classification} for a more general statement.
As a corollary, if an anomaly $\omega \in \rH^{3}(G;\U)$ can be realized on a spin chain with local Hilbert space dimension $n$, then there exists a sufficiently large $N \in \mathbb{N}^{*}$ such that
\[
[\omega^{\,n^{N}}] = [1] \in \rH^{3}(G;\U),
\]
that is, $\omega$ must have finite order in $\rH^{3}(G;\U)$.
In particular, since the generator of $\rH^{3}(\U;\U) \simeq \z$ has infinite order, such an anomaly cannot be realized in quantum spin systems. The relation between $\rH^{3}(G;\z[n^{-1}]/\z)$ and $\rH^{3}(G;\U)$ will be made more precise later in Sec.~\ref{sec:relate_coefficients}.

To recover the fore-mentioned classification in Ref.~\cite{Tu2025}, one needs to add all possible ancillas (i.e. stabilizing the spin systems), thus we have an inductive limit:
\beq
\lim_{n}\z[n^{-1}]/\z=\Q/\z
\eeq
Note that inductive limits commute with cohomology because it is an exact functor (see e.g. Prop.~5.33 of \cite{rotman2008introduction}), we have
\beq
\lim_{n}\rH^{*}(G;\z[n^{-1}]/\z)=\rH^{*}(G;\lim_{n}\z[n^{-1}]/\z)=\rH^{*}(G;\Q/\z)
\eeq
as proposed in Ref.~\cite{Tu2025}.

We comment that our result has implication not only for regularizing QFTs with anomalous symmetries, but also for condensed matter systems. Concretely,
\begin{question}
    Given a possibly non-onsite symmetry action, is there any easy way to decide whether it is anomalous or not, without calculating the anomaly cocycle explicitly?
\end{question}
This question is natural, since although the anomaly index admits an explicit algebraic expression (see Eq.~\eqref{eq:anomaly_index}), determining whether it represents a trivial cocycle is often not easy and may require substantial computational effort.
Moreover, when $G$ is infinite—for instance, when $G$ is a Lie group or contains translations—such checks become considerably more difficult and are often beyond the reach of brute-force computational methods.

Our theorem provides a simple criterion: if the order of an anomaly $\omega \in \rH^{3}(G;\U)$ is coprime to the local Hilbert space dimension $n$, then $\omega$ must be trivial.
As a concrete example, in the presence of time-reversal symmetry all anomalies have order $2$, and therefore no such anomaly can occur in a spin-$1$ chain (where $n=3$ and $\mathrm{gcd}(2,3)=1$).

The higher-form symmetry extension of our theorem also provides a useful guideline for constructing lattice models with anomalous higher-form symmetries, such as those associated with anyons.
In particular, if the anomaly—equivalently, the topological spin—has order coprime to the local Hilbert space dimension, then such a higher-form symmetry cannot be realized as an exact symmetry in spin models, even when non-onsite actions or tails are allowed. See \cite{liu2023symmetries,liu2026doeslatticehigherformsymmetry} the interplay of local dimension and the order of anomalies for on-site higher form symmetries.

Before turning to the proof, we remark that similar arguments have already been explored in 
Refs.~\cite{evington2022anomaloussymmetriesclassifiablecalgebras,
pacheco2025projectiverepresentationsoperatoralgebras} 
(see also \cite{izumi2024gkernelskirchbergalgebras}), 
where the authors considered the case in which $G$ is a discrete group%
\footnote{We thank Sergio Girón Pacheco for drawing these references to our attention.}. 
In the present work, we extend this line of reasoning to the setting of Lie groups. 
We also comment on the treatment of anti-unitary symmetries and higher-form symmetries 
in quantum spin systems.

\section{Algebras and symmetries}\label{sec:sym_action}
We work on infinite-volume spin system on $\z$. For each site $i\in\z$, we assign an $n\times n$ matrix algebra $\A_{i}\simeq\mathrm{M}_{n}(\bbC)$. Let $\Gamma\subseteq\z$ be a finite subset, the local operators supported on $\Gamma$ is defined to be
\beq
\A_{\Gamma}:=\bigotimes_{i\in\Gamma}\A_{i}
\eeq
The algebra of local operators is the given by the following colimit:
\beq
\A^{loc}:=\lim_{\Gamma\subseteq\z}\A_{\Gamma}
\eeq
We further complete $\A^{loc}$ with respect to the operator norm, obtaining the algebra of quasi-local operators, denoted by $\A^{ql}$, or simply $\A$.
Intuitively, $\A$ consists of operators with spatially decaying tails, with no restriction on the rate of decay.

The algebra $\A$ is a $C^*$-algebra, or more precisely, a uniformly hyperfinite (UHF) algebra; see Refs.~\cite{murphy2014c,bratteli2013operator1,bratteli2013operator2,Naaijkens_2017,Landsman:2017hpa,Liu2024LRLSM} for general background and further details.
In the present work, however, we will only rely on very little of $C^*$-algebraic techniques.

Recall that states are positive, normalized linear functionals on $\A$, meaning $\psi(a^{*}a)\geqslant0$ as well as $\psi(1)=1$. For us, it is important that since $\A$ is a UHF algebra, $\A$ admits a special state, called the tracial state (a.k.a. maximally mixed state or infinite temperature state) $\tau$ \cite{murphy2014c}. This state is uniquely characterized by $\tau(ab)=\tau(ba)$ for any $a,b\in\A$. Therefore, it is a normalized trace on $\A$. Later we will define a generalized determinant on $\A$ using the tracial state $\tau$ \cite{delaharpe2012fugledekadisondeterminantthemevariations}.

In quantum many-body physics, symmetry actions are required to preserve certain notion of locality. A popular choice is so-called quantum cellular automata (QCA) \cite{Else_2014,Tu2025}.
\begin{definition}\label{def:QCA}
    Let $\alpha\in\Aut(\A)$, one says $\alpha$ is a QCA if there exists $r_{\alpha}>0$  such that for any local operator $x\in \A_{X}$, we have $\alpha(x)\in\A_{B(X,r_{\alpha})}$, where $B(X,r_{\alpha}):=\{p\in \Lambda:\dist(p,X)\leqslant r_{\alpha}\}$. The constant $r_{\alpha}$ is called the radius of $\alpha$.
\end{definition}
QCA's are automorphisms which preserve the locality in a strict way: they map local operators to local operators with finite spreads of supports.
It is known that QCAs form a group, denoted by $\G^{\QCA}$. 
In one dimension, the structure of $\G^{\QCA}$ is well understood~\cite{Gross_2012}. 
In essence, one-dimensional QCAs can be decomposed into finite-depth quantum circuits composed with (partial) translations (see Refs.~\cite{arrighi2019overview,Farrelly_2020} for reviews).
Below, we outline the construction of the anomaly index associated with a symmetry action $\alpha: G \to \G^{\QCA}$ in spin chains.
For the construction of anomaly indices in higher dimensions, we refer the reader to Refs.~\cite{Else_2014,kapustin2025highersymmetriesanomaliescrossed,kawagoe2025anomalydiagnosissymmetryrestriction}.

Given an abstract symmetry group $G$, its symmetry action is a group homomorphism $\alpha: G \to \G^{\QCA}$, 
where, with a slight abuse of notation,\footnote{Earlier, the symbol $\alpha$ was used to denote an automorphism acting on operators. Here we instead use it to denote a homomorphism from the symmetry group $G$ to the group of quantum cellular automata $\G^{\QCA}$.}
$\alpha_{g}$ denotes the QCA implementing $g\in G$.

Such a symmetry action may include internal symmetries and/or lattice translations. 
The internal symmetry component, which is implemented by a finite-depth quantum circuit,  may act on-site or non-onsite.

First, suppose $\alpha$ is an internal symmetry action (\ie it contains no translation). For an arbitrary site, say, the origin, it can be shown that $\alpha$ can be decomposed as
\beq \label{eq: decomposition}
    \alpha=\alpha^{L} \, \alpha_0 \, \alpha^{R}
\eeq
where $\alpha^{R}$ (resp. $\alpha^{L}$) acts only on local operators supported on $[0, \infty)$ (resp. $(-\infty, 0)$), and $\alpha_0$ is the conjugation by a local unitary. Although $\alpha$ is a group homomorphism, in general $\alpha^{R}$ is not. In fact, for any $g, h\in G$,
\beq\label{eq:composition_half_chain}
\alpha^{R}_{g} \, \alpha^{R}_{h}=\Ad_{V_{g,h}} \, \alpha^{R}_{gh}
\eeq
where $V:G\times G\to \cU^{\ell}$ with $\cU^\ell$ the group of local unitaries is \textbf{not} necessarily a homomorphism, and $\Ad_V(a):=VaV^{*}$ for any $a\in\A$. The associativity of $\alpha^{R}$, \ie $\left( \alpha^{R}_{g} \, \alpha^{R}_{h} \right) \, \alpha^{R}_{k} = \alpha^{R}_{g} \, \left( \alpha^{R}_{h} \, \alpha^{R}_{k} \right)$ with $g, h, k\in G$, puts further constraints on $V$
\beq\label{eq:anomaly_index}
\Ad_{\omega_{g,h,k}}=1,\quad
    \omega_{g,h,k}
    :=V_{g,h}V_{gh,k}V_{g,hk}^{-1}\alpha^{R}_{g}(V_{h,k})^{-1}
\eeq
This implies that the operator $\omega$ is in fact a phase, since it commutes with all local operators.
One can verify that $\omega$ satisfies the $3$-cocycle condition \cite{Else_2014,kapustin2024anomalous}.
Moreover, redefining $V_{g,h}$ by a phase factor $\rho_{g,h}\in \U$ shifts $\omega$ by a $3$-coboundary $\delta\rho$.
Therefore, $\omega$ defines a cohomology class $[\omega]\in \rH^{3}(G;\U)$, which we refer to as the anomaly index associated with the symmetry action $\alpha$.
For a brief review of group cohomology, see Appendix~\ref{sec:group_cohomology}; see also Ref.~\cite{Else_2014} for a more detailed discussion.

If the symmetry action $\alpha$ contains lattice translations, one may stack the system with an additional copy on which the translation acts in the opposite direction~\cite{kapustin2024anomalous}.
The resulting symmetry action on the composite system, denoted by $\alpha_{\otimes}$, is free of translations.
We then define the anomaly index of $\alpha$ to be the anomaly index associated with $\alpha_{\otimes}$.

We now comment on several possible generalizations of the anomaly index.
First, one may allow symmetry actions with spatial tails by working with the notion of locality-preserving automorphisms introduced in Refs.~\cite{Ranard_2022,kapustin2024anomalous}.
More precisely, we adopt the following definition.

\begin{definition}\label{def:LPA}
An automorphism $\alpha \in \Aut(\A)$ is called a \emph{locality-preserving automorphism} (LPA) if there exists a nonnegative, monotonically decreasing function $f_{\alpha}(r)\searrow 0$ such that for any local operator $x \in \A_{X}$ and $r>0$, there exists an operator $x^{(r)} \in \A_{B(X,r)}$ satisfying
\begin{equation}
    \|\alpha(x) - x^{(r)}\| \leq f_{\alpha}(r)\,\|x\|.
\end{equation}
The function $f_{\alpha}$ is referred to as the \emph{tail} of $\alpha$.
The group of locality-preserving automorphisms is denoted by $\G^{lp}$.
\end{definition}

QCAs are clearly special cases of LPAs, corresponding to the strictly local situation in which the tail function satisfies
$f_{\alpha}(r)=0$ for all $r>r_{\alpha}$.
For this reason, we work exclusively with LPAs in the remainder of this paper.

When the symmetry group $G$ is a Lie group, it is natural to require the symmetry action to be smooth.
Since the group $\G^{lp}$ does not itself carry a canonical smooth structure, this requirement can be met by restricting to \emph{almost-local} LPAs, namely those whose tail functions satisfy
$f_{\alpha}(r)=O(r^{-\infty})$.
The subgroup of almost-local LPAs admits a natural smooth structure; see Ref.~\cite{kapustin2024anomalous} for a detailed discussion.
Throughout this work, whenever $G$ is a Lie group, we implicitly restrict to almost-local LPAs without further comment. We sketch the construction of anomaly index for Lie groups in Appendix \ref{sec:Proof_Lie}.

For anti-linear symmetries, one must instead work with \emph{twisted} LPAs, as explained in Ref.~\cite{liu2025twistedlocalitypreservingautomorphismsanomaly}.
The corresponding group of twisted LPAs is denoted by $\tilde{\G}^{lp}$.

In all of the above generalizations, the anomaly index can be defined in a similar manner.

\section{Warm up}
\subsection{Projective representations in $(0+1)d$}
To illustrate our main idea, we start with a projective representation, \ie 't Hooft anomalies in $(0+1)d$. Let $\rho$ be a projective representation on $\cH=\mathbb{C}^{n}$, \ie
\beq\label{eq:proj_rep}
\rho_{g}\rho_{h}=\omega_{g,h}\rho_{gh}
\eeq
for some phase $\omega\in\rH^{2}(G;\U)$. We take determinants on both sides of Eq.~\eqref{eq:proj_rep} and we then end up with:
\beq
\omega^{n}=\delta(\det\rho)\Rightarrow[\omega^{n}]=[1]\in\rH^{2}(G;\U)
\eeq
Therefore we know that given the Hilbert space $\cH=\bbC^{n}$, we have known that only $n$-torion subgroup of $\rH^{2}(G;\U)$ can be possibly realized on this Hilbert space. For example, a nontrivial projective reperesentation of $SO(3)$ or $\z_{2}\times\z_{2}$ cannot be realized on $\bbC^{n}$ if $n$ is odd. Conversely, any odd-dimensional representation of $SO(3)$ or $\z_{2}\times\z_{2}$ is automatically anomaly-free.

\subsection{Finite group symmetries and QCA actions}\label{sec:QCA}
In this subsection, we prove our main theorem in the case that $G$ is finite and the symmetry action $\alpha:G\to\G^{\QCA}$ has no tails.

In Eq.~\eqref{eq:anomaly_index}, since $V_{g,h}$ and $\,\alpha_{g}^{R}(V_{h,k})$ are local unitary operators for all $g,h,k\in G$, and $G$ is a finite group. It is hence possible to find a sufficiently large $N\in\mathbb{N}^{*}$, such that $V_{g,h},\, \alpha_{g}^{R}(V_{h,k})\in\A_{[0,N-1]}$ for all $g,h,k\in G$. Eq.~\eqref{eq:anomaly_index} is naturally an operator equation on $\cH_{[0,N-1]}:=\bigotimes_{i=0}^{N-1}\bbC^{n}$. Taking determinants for Eq.~\eqref{eq:anomaly_index} on $\cH_{[0,N-1]}$, we obtain
\beq\label{eq:ord_of_anomaly}
\omega^{n^{N}}=\delta(\det(V))^{-1}\Rightarrow [\omega]^{n^{N}}=[1]\in\rH^{3}(G;\U)
\eeq
i.e. $\omega$ has a finite order $n^{N}$. To derive the classification in Theorem~\ref{thm:main}, we fix the phases of $V_{g,h}$ by imposing the condition
\[
\det(V_{g,h})=1 .
\]
This choice can be achieved by shifting the cocycle $\omega$ with a $3$-coboundary in $C^{3}(G;\U)$.
With this “gauge fixing”, the determinant argument shows that $\omega$ takes values in $C^{3}\!\left(G;\,\z\!\left[n^{-1}\right]/\z\right)$ (since $N$ can be made arbitrarily large).

We emphasize, however, that the condition $\det(V_{g,h})=1$ does not completely fix the phase of $V_{g,h}$.
Indeed, for any $\lambda\in\U$ we have 
\beq
\det(\lambda V_{g,h})=\lambda^{\,n^{|V_{g,h}|}}\stackrel{!}{=}1\Rightarrow \lambda\in\z[n^{-1}]/\z
\eeq
where $|V_{g,h}|$ denotes the size of the support of $V_{g,h}$.
As a result, there remains a residual phase ambiguity $\lambda$ in the choice of $V_{g,h}$,
\beq
\lambda \in C^{2}(G;\,\z[n^{-1}]/\z)
\eeq

Thus, only the cohomology class of $\omega\in\rH^{3}(G;\z[n^{-1}]/\z)$ is well-defined, as we claimed in Theorem \ref{thm:main}.

As a quick corollary, if the order of $\omega$ in $\rH^{3}(G;\U)$ is coprime to $n$, then $[\omega^{n^{N}}]=[1]$ implies $[\omega]=[1]$ by elementary number theory.

\section{The proof to the main theorem}
\subsection{de la Harpe-Skandanis determinant}
The same idea in Sec.~\ref{sec:QCA} generalizes to Lie group $G$ as well as LPA symmetry actions (i.e. with tails), with the help of de la Harpe-Skandanis determinants in operator algebras \cite{delaharpe2012fugledekadisondeterminantthemevariations}. We call it the (generalized) determinant for simplicity and sketch the main idea below, see Ref.~\cite{delaharpe2012fugledekadisondeterminantthemevariations} for more details.

We begin by recalling a determinant formula for unitary matrices.
Given $u \in U(n)$, a classical result states that
\beq\label{eq:det_log}
\det(u)=\exp\!\bigl(\mathrm{Tr}(\log u)\bigr),
\eeq
which can be established, for example, by diagonalizing $u$.

A more geometric formulation of Eq.~\eqref{eq:det_log} proceeds as follows.
Choose a smooth path $\gamma:[0,1]\to U(n)$ such that $\gamma(0)=\id$ and $\gamma(1)=u$, and define
\begin{equation}\label{eq:dlHS_det}
\Delta(u):=\frac{1}{2\pi \rmi}\int_{\gamma}\mathrm{Tr}\bigl(g^{-1}\rd g\bigr)
\quad \mathrm{mod}\ \z .
\end{equation}
where $g^{-1}\rd g$ is the Maurer-Cartan 1-form on $U(n)$.
It is straightforward to verify that
\[
\det(u)=\exp\!\bigl(2\pi i\,\Delta(u)\bigr).
\]

The reduction modulo $\z$ in Eq.~\eqref{eq:dlHS_det} is necessary because $\pi_{1}(U(n))\simeq \z$, so the integral generally depends on the choice of the path $\gamma$.
However, different choices of $\gamma$ can change $\Delta(u)$ only by an integer, and hence the fractional part of $\Delta(u)$ is well defined.

Eq.~\eqref{eq:dlHS_det} can be generalized to the unitary group in $\A$ in a straightforward way. Let us recall that the unitary group in $\A$ means
\beq
\cU(\A):=\{u\in\A:u^{*}u=1\}
\eeq
Rcall that on $\A$, there is a tracial state $\tau$, which serves as a normalized trace on $\A$. One can thus define
\beq
\Delta_{\tau}(u):=\frac{1}{2\pi \rmi}\int_{\gamma}\tau(g^{-1}\rd g)\quad\mathrm{mod}\ \z[n^{-1}] 
\eeq
The reduction by modding $\z[n^{-1}]$ is again from the path dependence, and the fractional part comes from the normalization of $\tau$.

And the generalized determinant is defined as
\beq
\mathrm{det}_{\tau}:\cU(\A)\to \frac{\U}{\z[n^{-1}]/\z}, \quad u\to \exp(2\pi i\Delta_{\tau}(u))
\eeq
Using the left invariance of $g^{-1}\rd g$, it is easy to see $\det_{\tau}$ is a group homomorphism. Besides, since $\tau\circ\beta=\tau$ for any $\beta\in\Aut(\A)$, it is easy to see
\beq
\mathrm{det}_{\tau}(\beta(u))=\mathrm{det}_{\tau}(u),\quad\forall u\in\A,\,\beta\in\Aut(\A)
\eeq
Let us now define the special unitary group of $\A$:
\beq
\cSU(\A):=\{u\in\cU(\A): \Delta_{\tau}(u)=0,\quad\mathrm{mod}\ \z [n^{-1}] \}
\eeq

In the case that $\alpha:G\to\G^{lp}$, the strategy of proof in Sec.~\ref{sec:QCA} generalizes in a straightforward way. We make a gauge-fixing so that 
\beq
\Delta_{\tau}(V_{g,h})=0,\quad\mathrm{mod}\ \z [n^{-1}]
\eeq
With this gauge-fixing condition, it is straightforward to see that the resulting classification is again given by $\rH^{3}(G;\z[n^{-1}]/\z)$. This completes the proof of our main theorem in the case where $G$ is discrete and the symmetry action is implemented by $\G^{lp}$.

\subsection{Comment on Lie group case}
As discussed in \cite{kapustin2024anomalous}, the anomaly index for a Lie group $G$ can be described using so-called differentiable group cohomology, which we briefly review in Appendix~\ref{sec:differentiable_group_cohomology}. The basic idea is to choose a good simplicial covering of $\{G^{\times p}\}_{p\in\mathbb{N}}$, where $G^{\times p}$ denotes the $p$-fold Cartesian product of $G$, and to define $\alpha_g^{R}$ and $V_{g,h}$ locally on each patch so that the smoothness can be retained. One then shows that these local data can be glued together by suitable transition functions; see Section~A.6.2 of \cite{kapustin2024anomalous} for details.

As in the previous cases, one can further impose a gauge-fixing condition so that both $V_{g,h}$ and the transition functions lie in the special unitary group $\cSU(\A)$. With this choice, the resulting classification again reduces to $\rH^{3}_{\diff}(G;\z[n^{-1}]/\z)\simeq\rH^{3}(G;\z[n^{-1}]/\z)$, where we have used Prop.~\ref{prop:discrete} and the fact that $\z[n^{-1}]/\z$ is discrete.

As a corollary, since $\rH^{3}(\U;\z[n^{-1}]/\z)=0$ for any $n$, the $\U$ symmetry on a spin chain is always non-anomalous. This recovers Proposition A.3 of \cite{kapustin2024anomalous}.

We give the detailed proof to this case in Appendix \ref{sec:Proof_Lie}.

\section{The coefficient and operator $K$-theory}
\subsection{Relating different coefficients}\label{sec:relate_coefficients}
Here, we give a more explicit relation between $\rH^{*}(G;\z[n^{-1}]/\z)$, $\rH^{*}(G;\Q/\z)$ and $\rH^{*}(G;\U)$.
First, the following short exact sequence
\beq
1\to\z\hookrightarrow\z[n^{-1}]\twoheadrightarrow\z[n^{-1}]/\z\to 1
\eeq
induces a long exact sequence:
\beq
\dots\to\rH^{3}(G;\z[n^{-1}])\to\rH^{3}(G;\z[n^{-1}]/\z)\to\rH^{4}(G;\z)\to\rH^{4}(G;\z[n^{-1}])\to\dots
\eeq
Besides, we have\footnote{This is ensured by $\rH^{*}(G;\R)=0$ for $*>0$.} $\rH^{3}(G;\U)\simeq\rH^{4}(G;\z)$. Thus, the map $\rH^{3}(G;\z[n^{-1}]/\z)\to\rH^{4}(G;\z)\simeq\rH^{3}(G;\U)$ is neither injective nor surjective in general. Besides, since localization is an exact functor (see e.g. Corollary 4.81 of \cite{rotman2008introduction}), $P_{n}^{-1}$ commutes with the cohomology:
\beq\label{eq:localization_cohomology}
\rH^{*}(G;\z[n^{-1}])\simeq P_{n}^{-1}\rH^{*}(G;\z)
\eeq
The image $\im(\rH^{3}(G;\z[n^{-1}]/\z)\to\rH^{3}(G;\U)$ is nothing but the kernel of above localization map Eq.~\eqref{eq:localization_cohomology}. By exactness, one recovers the earlier condition Eq.~\eqref{eq:ord_of_anomaly} from the kernel of the localization map $\rH^{3}(G;\z)\to P_{n}^{-1}\rH^{3}(G;\z)$.

If one allows stabilization—namely, adjoining infinitely many ancillas, or equivalently passing to the inductive limit in $n$—the situation simplifies considerably. 
The localization map
\beq
\rH^{4}(G;\z)\longrightarrow\rH^{4}(G;\Q)
\eeq
annihilates the torsion subgroup, while the free part survives. 
Consequently, $\rH^{3}(G;\Q/\z)$ can be identified with the torsion subgroup of $\rH^{3}(G;\U)$.

\subsection{Connection to operator $K$-theory}
One may naturally wonder what the meaning of the coefficient $\z[n^{-1}]/\z$ is. From \cite{delaharpe2012fugledekadisondeterminantthemevariations}, the codomain of the generalized determinant on the unitary group is actually given by $\U/\tilde{K}_{0}(\A)$, where $\tilde{K}_{0}(\A)$ is the reduced Grothendieck group of $\A$, see \cite{Blackadar:2006OA,murphy2014c} for the detailed definition. Indeed, as is shown in V.1.1.16 of Ref.~\cite{Blackadar:2006OA}, when $\A=\ovl{\bigotimes_{i\in\z}\mathrm{M}_{n}(\bbC)}^{\|\cdot\|}$ (as we defined earlier), we have $K_{0}(\A)=\z[n^{-1}]$ and hence $\tilde{K}_{0}(\A)=\z[n^{-1}]/\z$. Actually, the relevance of the $K_{0}$ group here is not completely surprising. Since the definition of generalized determinants is clearly relevant to $\pi_{1}(\cU(\A))$, which can be related to $K_{0}(\A)$ by Bott periodicity (at least after stabilization), see \cite{Blackadar:2006OA}.

Therefore, the general case of the classification can be summarized as
\begin{theorem}\label{thm: classification}
    Let $\alpha:G\to\G^{lp}$ be a symmetry action on a $(1+1)d$ spin chain whose quasi-local algebra is $\A$ (the local dimension is not assumed to be the same everywhere), then the anomaly index $\omega$ of $\alpha$ satisfies:
    \beq
    [\omega]\in\rH^{3}(G;\tilde{K}_{0}(\A))
    \eeq
\end{theorem}
For this general case, $\tilde{K}_{0}(\A)$ is simply obtained by inverting all local dimensions in the spin chain and modding out the integer part.

 To recover the classification in Ref.~\cite{Long_2025, Tu2025}, we add all possible ancillas to stabilize the system. Therefore $\A$ becomes the universal UHF algebra, whose $K_{0}$ becomes $\Q$ since we have inverted all primes.

We emphasize that $\tilde{K}_{0}(\A)$ is \emph{functorial} in $\A$. 
Let $\cB$ be the quasi-local algebra of another spin system. 
Any morphism $\A \to \cB$—for instance an algebra homomorphism or an $\A$–$\cB$ bimodule—induces a map 
\[
\tilde{K}_{0}(\A)\longrightarrow \tilde{K}_{0}(\cB).
\]
Consequently, there is an induced homomorphism on anomaly indices
\[
\rH^{*}(G;\tilde{K}_{0}(\A))
\longrightarrow
\rH^{*}(G;\tilde{K}_{0}(\cB)).
\]
At present, however, the precise physical interpretation and implications of this functoriality remain unclear to us.

\section{Generality of our proof}
We expect that our strategy applies whenever the anomaly index is \emph{locally computable}, namely when it can be expressed as a product of quasi-local unitary operators (at least on each open set of a covering of $\{G^{\times p}\}_{p\in\mathbb{N}}$). In this situation, one can always perform a gauge fixing such that all these unitary operators have trivial (generalized) determinants.

As a non-example, the symmetry index associated with symmetry-protected topological
(SPT) phases is \textbf{not} locally computable in our sense.
Indeed, its definition crucially depends on the global structure of the underlying quantum
state; see
\cite{Ogata2019split,Ogata2021,Sopenko_2021,Ogata_2022,ogata20222dfermionicsptcrt,sopenko2025reflectionpositivityrefinedindex}
for further details.

Consequently, in the classification of bulk SPT phases, the relevant coefficient group
is $\U$, rather than $\tilde{K}_{0}(\A)$ as in Theorem~\ref{thm: classification}.
This mismatch implies that a naïve bulk--boundary correspondence may fail.
In particular, there exist bulk SPT phases in $(2+1)d$ spin systems whose boundary anomalies
cannot be realized in a $(1+1)d$ spin chain, such as Chern insulators and
their analogues in spin systems
\cite{Qi_2011,He_2015,shankar2018topologicalinsulatorsreview}.
Ref.~\cite{Tu2025} proposes a refined version of the bulk–boundary correspondence designed to remedy this mismatch.

This local computability is known to hold for $(2+1)$-dimensional systems with ordinary symmetries \cite{kapustin2025highersymmetriesanomaliesquantum,kawagoe2025anomalydiagnosissymmetryrestriction}, as well as for higher-form symmetries \cite{kapustin2025highersymmetriesanomaliescrossed,feng2025onsiteabilityhigherformsymmetries,Feng_2026,Kobayashi_2026}, Ref.~\cite{Tu2025,czajka2025anomalieslatticehomotopyquantum} also provides an systematic way to write down anomaly indices in terms of local unitary operators\footnote{All these works assume QCA symmetry actions, i.e.\ symmetry actions without tails. We expect that their formulas for anomaly indices remain valid even when symmetry actions with tails are allowed.}.
We further note that the same strategy readily extends to anti-unitary symmetries as well as fermionic systems.

\section{Discussion}
We present a unified algebraic perspective on Nielsen–Ninomiya–type no-go theorems in quantum spin systems associated with group-cohomological anomalies. Our approach departs from the conventional analytic treatments and instead emphasizes the role of algebraic structures associated with symmetry actions, anomaly indices and the local computability of them.

A central message of this work is that the obstruction underlying Nielsen--Ninomiya type theorems is not specific to free theories or to particular lattice realizations, but is instead a manifestation of a more general algebraic incompatibility between the dimension of local Hilbert spaces and anomaly data. From this viewpoint, the traditional Nielsen--Ninomiya theorem appears as a special case of a broader principle governing the realizability of anomalous symmetry actions in lattice spin systems.

Our method highlights the importance of the \emph{local computability} of anomaly indices. Whenever the anomaly index can be expressed locally in terms of quasi-local unitary operators, one can perform a suitable gauge fixing that trivializes their (generalized) determinants, leading to strong constraints on the existence of lattice regularizations. This criterion provides a unifying explanation for a class of no-go results, encompassing ordinary internal symmetries in $(2+1)$ dimensions as well as higher-form symmetries.

Several open questions remain. First, while our analysis assumes that the anomaly index is locally computable, it would be highly desirable to develop a systematic method for constructing explicit expressions for such indices. For QCA symmetry actions, systematic constructions of anomaly indices have been explored in Refs.~\cite{Tu2025,czajka2025anomalieslatticehomotopyquantum}, based on the $\Omega$-spectrum of QCAs. It would therefore be interesting to generalize these constructions to settings that incorporate symmetry actions with tails.

Second, our discussion has largely been restricted to exact symmetries. The present results do not exclude the possibility of realizing such anomalies on spin systems with emergent or emanant anomalies \cite{Metlitski2018, Cheng2022,Seiberg_2024}. Also, one can also avoid our no-go theorem by considering infinite-dimensional Hilbert spaces (i.e. lattice bosons or rotors) as well as non-tensor-factorized Hilbert spaces (e.g. by imposing Gauss law constraints).

Finally, it is an interesting open problem to determine whether our approach can be extended to systems with non-invertible symmetries, where the standard group-cohomological description of anomalies is no longer available and one has to consider so-called fiber functors instead \cite{thorngren2019fusioncategorysymmetryi,thorngren2021fusioncategorysymmetryii,evans2025operatoralgebraicapproachfusion}.

We hope that the perspective developed here helps clarify the conceptual content of Nielsen--Ninomiya type theorems and provides a useful framework for understanding and extending no-go results across different symmetry types and spatial dimensions.

\section{Acknowledgements}
The author would like to thank Dominic Else, Zhi Li, Han Ma, Xinping Yang, Jinmin Yi and Liujun Zou for inspiring discussions. Research at Perimeter Institute is supported in part by the Government of Canada
through the Department of Innovation, Science and Industry Canada and by the Province of Ontario through
the Ministry of Colleges and Universities. The author is also supported by the Simons Collaboration on Global Categorical Symmetries through Simons Foundation grant 888996.

\appendix
\section{Group cohomology and differentiable group cohomology}\label{sec:group_cohomology}

In this section, we review the basics of group cohomology and differentiable group cohomology. For group cohomology, there are many materials in the literature \cite{brown2012cohomology,Weibel_1994_group,Chen2010,Yang_2017,kobayashi2025projectiverepresentationsbogomolovmultiplier}. For differentiable group cohomology, see appendix A.1 of Ref. \cite{kapustin2024anomalous} and Ref. \cite{brylinski2000differentiable}. We will only cover the motivations and basics here.

\subsection{Projective representations in quantum mechanics}\label{sec:proj_rep}

To motivate group cohomology, let us begin with projective representations in quantum mechanics.
Consider a (for simplicity, discrete and unitary) symmetry group $G$ acting on a Hilbert space $\cH$.
In the standard setting, a symmetry action is described by a homomorphism
\[
\rho: G \to U(\cH),
\]
that is, a unitary representation satisfying
\[
\rho(g)\rho(h)=\rho(gh), \qquad \forall\, g,h\in G .
\]

However, in quantum mechanics, physical states are not vectors in $\cH$ but rather \emph{rays}:
the vectors $|\psi\rangle$ and $e^{i\theta}|\psi\rangle$ represent the same physical state.
Accordingly, the true space of states is the projective Hilbert space $P(\cH)$, rather than $\cH$ itself.
This observation allows for a more general notion of symmetry action, in which the group law is respected
only up to a phase:
\[
\rho(g)\rho(h)=\omega(g,h)\,\rho(gh),
\]
where $\omega(g,h)\in \U$.\footnote{Strictly speaking, one must show that the phase $\omega(g,h)$
is independent of the state on which the operators act. This follows from the coherence of quantum
states; see Sec.~2.2 of Ref.~\cite{weinberg2005quantum} for a detailed discussion.}
Such a map $\rho$ is called a \emph{projective representation} of $G$, and $\omega$ is referred to as
its \emph{factor set}.

Since operator multiplication is associative, we must have
\[
(\rho(g)\rho(h))\rho(k)=\rho(g)(\rho(h)\rho(k)), \qquad \forall\, g,h,k\in G .
\]
Substituting the projective composition law, this condition imposes a constraint on $\omega$,
which will lead naturally to the notion of a group $2$-cocycle.

This imposes the following constraint on $\omega$,
\beq\label{eq:2-cocycle_condition}
\omega(g,h)\omega(gh,k)=\omega(g,hk)\omega(h,k)
\eeq
Any function $G\times G\to \U$ satisfying Eq. \eqref{eq:2-cocycle_condition} is called a 2-cocycle. Furthermore, one can redefine the phase of $\rho(g)\to \tilde{\rho}(g)=\rho(g)\eta(g),\eta(g)\in\U$ (we do not require $\eta:G\to \U$ to be a homomorphism), and the resulting 2-cocycle is
\beq\label{eq:shift_2-coboundary}
\tilde{\omega}(g,h)=\omega(g,h)\eta(g)\eta(h)\eta(gh)^{-1}
\eeq
One can easily check that $\tilde{\omega}$ again satisfies the 2-cocycle condition, Eq. \eqref{eq:2-cocycle_condition}. If there exists $\eta(g)$ such that $\tilde{\omega}(g,h)=1$ for all $g,h\in G$, then we say that $\omega$ is a 2-coboundary or trivial. Any two 2-cocycles $\omega$ and $\tilde{\omega}$ related by Eq. \eqref{eq:shift_2-coboundary} are viewed as equivalent, since they differ only by the artificial choice of phase factors $\eta(g)$ of representation matrix $\rho(g)$. We write $\omega\sim \tilde{\omega}$ if $\omega$ and $\tilde\omega$ are equivalent. The space of 2-cocycles modulo this equivalence $\sim$ is the so-called the degree 2 group cohomology of $G$, denoted by $\rH^{2}(G;\U)$.

\subsection{Group cohomology}

We now recall the general definition of group cohomology.
Let $G$ be a discrete group.
The classifying space $BG$ can be modeled simplicially by the nerve of $G$ \cite{Weibel_1994_simplicial},
whose set of $n$-simplices is given by
\[
(BG)_n = G^n , \qquad n=0,1,2,\dots .
\]
This simplicial structure is equipped with face maps
$d_k : G^n \to G^{n-1}$ for $k=0,1,\dots,n$, defined explicitly by
\beq\label{eq:face_maps}
d_{k}(g_{1},\dots,g_{n})=
\begin{cases}
(g_2,\dots,g_n), & k=0,\\
(g_1,\dots,g_k g_{k+1},\dots,g_n), & 0<k<n,\\
(g_1,\dots,g_{n-1}), & k=n .
\end{cases}
\eeq
These maps satisfy the simplicial identities, which imply that the alternating sum
\[\label{eq:group_coboundary}
d := \sum_{k=0}^{n} (-1)^k d_k
\]
obeys $d^2=0$.

Let $A$ be an abelian group, which we always equip with the discrete topology.
Typical examples include $\z_2$, $\z$, $\z[n^{-1}]/\z$, and $\U$.
Throughout this appendix, we write the group law of $A$ additively,
even when $A=\U$; this differs from the multiplicative convention used
in the main text.

The group of $A$-valued $n$-cochains on $G$ is defined as
\[
C^n(G,A) := \mathrm{Map}(G^n,A) ,
\]
and we denote the direct sum over all degrees by
$C^\bullet(G,A)=\bigoplus_{n\ge 0} C^n(G,A)$.
For instance, an element $\lambda\in C^2(G,A)$ is simply a function
$\lambda:G^2\to A$.

Given $\omega\in C^{n-1}(G,A)$, each face map $d_k:G^n\to G^{n-1}$
induces a pullback
\[
d_k^*\omega := \omega\circ d_k \in C^n(G,A) .
\]
The coboundary operator $\delta:C^{n-1}(G,A)\to C^n(G,A)$ is then defined by
\[
(\delta\omega)(g_1,\dots,g_n)
:= \sum_{k=0}^{n} (-1)^k\, \omega\bigl(d_k(g_1,\dots,g_n)\bigr) .
\]
By construction, one has $\delta^2=0$, so that
$\bigl(C^\bullet(G,A),\delta\bigr)$ forms a cochain complex.

\begin{definition}\label{def:group_cohomology}
    An $n$-cochain $\omega:G^{n}\to A$ is said to be an $n$-cocycle if $\delta \omega=0$. We denote the space of all $n$-cocycles by $\mathrm{Z}^{n}(G;A)$. Besides, if an $n$-cocycle $\omega$ satisfies $\omega=\delta \eta$ for some $\eta\in C^{n-1}(G;A)$, it is called an $n$-coboundary. The space of all $n$-coboundary is denoted as $\mathrm{B}^{n}(G;A),n>1$. Besides, $\mathrm{B}^{1}(G;A)$ is defined to be 0.
\end{definition}

\begin{definition}
The degree-$n$ group cohomology of a discrete group $G$ with coefficients in an
abelian group $A$ is defined as
\beq\label{eq:group_coho}
\rH^{n}(G;A)
:= \frac{\mathrm{Z}^{n}(G;A)}{\mathrm{B}^{n}(G;A)} ,
\eeq
where $\mathrm{Z}^{n}(G;A)=\ker\delta$ denotes the group of $n$-cocycles and
$\mathrm{B}^{n}(G;A)=\mathrm{im}\,\delta$ denotes the group of $n$-coboundaries.

Equivalently, $\rH^{n}(G;A)$ consists of equivalence classes of $n$-cocycles
$\omega\in \mathrm{Z}^{n}(G;A)$ under the equivalence relation
\[
\omega \sim \omega + \delta\eta ,
\qquad \eta\in C^{n-1}(G;A) .
\]
\end{definition}

\begin{example}
Let $\omega:G\to A$ be a $1$-cochain, where $G$ acts trivially on $A$.
The coboundary $\delta\omega\in C^{2}(G,A)$ is given by
\beq
\delta\omega(g_{1},g_{2})
= (d_{0}^{*}\omega - d_{1}^{*}\omega + d_{2}^{*}\omega)(g_{1},g_{2})
= \omega(g_{1}) + \omega(g_{2}) - \omega(g_{1}g_{2}) ,
\eeq
where we have used the face maps in Eq.~\eqref{eq:face_maps}. For instance,
\beq
d_{1}^{*}\omega(g_{1},g_{2})
= \omega\bigl(d_{1}(g_{1},g_{2})\bigr)
= \omega(g_{1}g_{2}) .
\eeq
Thus, $\omega$ is a $1$-cocycle if and only if
\[
\omega(g_{1}g_{2}) = \omega(g_{1}) + \omega(g_{2}) ,
\]
namely, $\omega$ is a group homomorphism. It follows that
\beq
\rH^{1}(G;A) \cong \rHom(G,A) .
\eeq
\end{example}

\begin{example}
Now consider a $2$-cochain $\omega:G^{2}\to A$.
A direct computation shows that
\beq
\delta\omega(g_{1},g_{2},g_{3})
= \omega(g_{2},g_{3})
- \omega(g_{1}g_{2},g_{3})
+ \omega(g_{1},g_{2}g_{3})
- \omega(g_{1},g_{2}) .
\eeq
If one writes the group law of $A$ multiplicatively rather than additively,
the cocycle condition $\delta\omega=0$ is precisely the $2$-cocycle condition
appearing in projective representations, cf.\ Eq.~\eqref{eq:2-cocycle_condition}.

Two $2$-cocycles differing by a coboundary $\delta\eta$, with $\eta\in C^{1}(G,A)$,
are equivalent. Explicitly,
\beq
\omega(g_{1},g_{2})
\;\longmapsto\;
\tilde{\omega}(g_{1},g_{2})
= \omega(g_{1},g_{2})
+ \eta(g_{1})
+ \eta(g_{2})
- \eta(g_{1}g_{2}) .
\eeq
In the language of projective representations, this corresponds to redefining
the representation matrices by a phase, as in Eq.~\eqref{eq:shift_2-coboundary}.
\end{example}

Group cohomology in higher degrees plays a central role in the classification
of \,'t~Hooft anomalies in quantum many-body systems.
We will discuss this connection in more detail in
Sec.~\ref{sec:sym_action}.

\begin{remark}
The geometric content behind Eqs.~\eqref{eq:face_maps} and~\eqref{eq:group_coho}
is that group cohomology can be identified with the simplicial cohomology
of $BG$.
We refer to Ref.~\cite{Weibel_1994_simplicial} for a systematic introduction.
\end{remark}

\subsection{Differentiable group cohomology}\label{sec:differentiable_group_cohomology}

It is natural to attempt to extend the above definition of group cohomology
from discrete groups to Lie groups.
A naïve approach would be to require group cochains to be smooth,
\ie one considers smooth maps
\[
\omega : G^{n} \longrightarrow A ,
\]
where $G$ is a Lie group and $A$ is an abelian Lie group on which $G$ acts smoothly.
We denote the resulting cohomology of smooth cochains by
$\rH^{*}_{s}(G;A)$.

However, this definition turns out to be too restrictive.
By a theorem of van Est (see Ref.~\cite{Stasheff1978ContinuousCO} for a proof),
if $G$ is a connected compact Lie group, then
\beq
\rH^{n}_{s}(G;A)=0 ,
\qquad n>0 .
\eeq
More generally, if $G$ is compact but not necessarily connected, one has
\beq
\rH^{n}_{s}(G;A)\simeq \rH^{n}\bigl(\pi_{0}(G);A\bigr),
\qquad n\ge 0 .
\eeq
As a consequence, this naïve smooth group cohomology fails to encode
any nontrivial information about the smooth structure of $G$.

There are, however, several well-established ways to define
cohomology theories for Lie groups that remain nontrivial and useful;
see Ref.~\cite{Wagemann2011cohomology} for a systematic comparison.
Roughly speaking, these approaches fall into three classes:
\begin{enumerate}
    \item \textbf{Measurable cochains.}
    One replaces smooth cochains by measurable ones.
    The resulting theory is commonly referred to as
    \emph{Borel group cohomology} in the physics literature%
    \footnote{This notion should not be confused with Borel
    equivariant cohomology, which is often called ``Borel cohomology''
    in the mathematical literature.}.
    We denote it by $\rH^{*}_{\mathrm{B}}(G;A)$
    and refer to Refs.~\cite{Chen2010,Ogata_2021} for physical applications.

    \item \textbf{Locally smooth cochains.}
    One weakens the smoothness requirement by demanding that
    $\omega:G^{n}\to A$ be smooth only in a neighborhood of the identity
    $(1,1,\dots,1)\in G^{n}$.
    The corresponding cohomology theory is denoted by
    $\rH^{*}_{\mathrm{loc},s}(G;A)$.

    \item \textbf{Simplicial cochains.}
    In this approach, one models the classifying space $BG$
    using a simplicial manifold and defines cochains as smooth
    functions on local charts.
    These local data are glued together using transition functions
    on overlaps.
    The resulting cohomology theory is usually denoted by
    $\rH^{*}_{\diff}(G;A)$,
    which coincides with the simplicial smooth cohomology
    $\rH^{*}_{\mathrm{simp},s}(G;A)$ in the notation of
    Ref.~\cite{Wagemann2011cohomology}.
\end{enumerate}
Conceptually, the discontinuities in the first two constructions reflect coordinate singularities rather than genuine physical obstructions. 
Smooth cochains or liftings should be defined on local charts of $G$, forming part of an atlas, instead of being required to exist globally. 
Any apparent discontinuity arises only when attempting to extend these local descriptions beyond their natural domains.

We give more details on the construction of $\rH^{*}_{\diff}(G;\U)$ since we will need it in Appendix \ref{sec:Proof_Lie}. Our notations and presentations closely follow Ref.~\cite{brylinski2000differentiable, kapustin2024anomalous}.

Let $\{U_{p}^{(a)}\}_{a\in I_{p}}$ be an open simplicial covering of $G^{\times p}$, where $I_{p}$ is an indexing simplicial set. We require this open covering to be compatible with the face maps Eq.~\eqref{eq:face_maps}: $d_{k}U_{p}^{(a)}\subseteq U_{p-1}^{(d_{k}a)}$; such simplicial covering is known to exist in Ref.~\cite{Brylinski1994}. As explained above, we will be interested in the intersections of these open coverings. Therefore, we define the disjoint union of $q$-fold intersections (i.e. intersections between $(q+1)$ covering open sets) of these covering open sets:
\beq
U_{p,q}:=\bigsqcup_{a_{0},a_{1},\dots,a_{q}}U_{p}^{(a_{0})}\cap U_{p}^{(a_{1})}\cap \dots\cap U_{p}^{(a_{q})}
\eeq
For example, $U_{p,0}$ is just the covering open sets of $G^{\times p}$; $U_{p,1}$ is all possible intersections between them; $U_{p,2}$ is all possible 2-fold intersections and so on.
Given an abelian Lie group $A$, we define
\beq\label{eq:double_complex}
C^{p,q}(U_{\bullet,\bullet},\underline{A}):=\bigoplus_{a_{0},\dots,a_{q}\in I_{p}}C^{\infty}(U_{p}^{(a_{0})}\cap \dots\cap U_{p}^{(a_{q})},A)
\eeq
Besides the coboundary operator Eq.~\eqref{eq:group_coboundary}, there is another differential operator on $C^{p,q}(U_{\bullet,\bullet},\underline{A})$ into a \textit{double} complex. Define $\partial_{k}:C^{p,q}\to C^{p,q+1}$ such that it omits $a_{k}$ in each summand of Eq.~\eqref{eq:double_complex}. The differential $\partial$ is defined\footnote{In \cite{kapustin2024anomalous} $\partial$ is written as $\delta$, which is reserved for group coboundary operator in our work.} as:
\beq
\partial :=\sum_{k=0}(-1)^{k}\partial_{k}
\eeq
For example, let $f\in C^{1,0}(U_{\bullet,\bullet},A)$, i.e. $f$ is a collection of $A$-valued smooth functions $\{f_{i}\}_{i\in I_{1}}$, then $\partial$ acts as:
\beq
\partial f|_{U_{i}\cap U_{j}}=f_{i}-f_{j}
\eeq
It is easy to check $\partial^{2}=0$ and $\partial \circ\delta=\delta\circ\partial$. Therefore, $D:=\delta+\partial$ defines a total differential operator for this double complex $C^{p,q}$. The differentiable group cohomology $\rH^{*}_{\diff}(G;A)$ is defined to be the (hyper-)cohomology of this double complex.

In order to get rid of the dependence on the choice of covering sets, one can work with \textit{good} coverings. This means each $\{U_{a_{k}}\}_{a_{k}\in I_{1}}$ as well as their $q$-fold intersections are all contractible. Such good coverings are known to exist in Ref.~\cite{Brylinski1994}.

\begin{example}
Let us start with an almost trivial example, $\rH^{1}(G;A)$ where $G$ acts trivially on $A$. In this case, we have a collection $\{\omega_{i}\}_{i\in I_{1}}$ be smooth $A$-valued functions defined on each $U_{i},i\in I_{1}$ as well as smooth 0-cochains $\eta$, we have
\beq
\begin{split}
    \delta\omega&=1\\
    \partial\omega&=\delta\eta\\
    \partial\eta&=1
\end{split}
\eeq
The first equation means $\omega(g)\omega(h)=\omega(gh)$, where we write $\omega(g):=\omega_{i}(g)$ if $g\in U_{i}$. The consistency of this notation is due to the second equation: it means on $U_{i}\cap U_{j}$, we have $\omega_{i}(g)\omega_{j}(g)^{-1}=(\delta\eta)(g)=1$ since $\eta$ is a 0-cochain and thus $\{\omega_{i}\}_{i\in I_{1}}$ can be patched together directly. The last equation simply says $\eta$ is automatically a global function on $G$. Thus, we find that
\beq
\rH^{1}(G;A)=\mathrm{Hom}_{\mathrm{smooth}}(G;A)
\eeq
that is, the smooth homomorphisms from $G$ to $A$.

\end{example}

\begin{example}
We give a more interesting example, i.e. the projective representation of Lie group $G$ on a Hilbert space $\cH$, i.e.
\beq
\rho:G\to \mathrm{PU}(\cH):=\mathrm{U}(\cH)/\U
\eeq
For each $U^{(a)}_{1}\in U_{1,0}=\{U^{(a)}_{1}\}_{a\in I_{1}}$, one can lift $\rho$ to $\tilde{\rho}_{a}:U_{1}^{(a)}\to \mathrm{U}(\cH)$. Similarly, for any $U^{(b)}_{2}\in U_{2,0}$ and $(g,h)\in U^{(b)}_{2}$, we have
\beq\label{eq:local_2_cocycle}
\tilde{\rho}_{d_{2}(b)}(g)\tilde{\rho}_{d_{0}(b)}(h)=\omega_{b}(g,h) \tilde{\rho}_{d_{1}(b)}(gh),\quad \omega_{b}(g,h)\in\U
\eeq
where we have used that the indexing simplicial set of $U_{\bullet,\bullet}$ is compatible with face maps Eq.~\eqref{eq:face_maps}. It is easy to verify that $\delta\omega=1$ as usual. There are also transition functions: on $U_{1,1}:=\bigsqcup_{a_{0},a_{1}\in I_{1}}U_{1}^{(a_{0})}\cap U_{1}^{(a_{1})}$, we define
\beq\label{eq:transition_func}
\eta_{a_{0},a_{1}}(g):=\tilde{\rho}_{a_{0}}(g)\tilde{\rho}_{a_{1}}(g)^{-1}\in \U
\eeq
They come from the ambiguity for the choice when lifting $\rho(g)$. When $g\in U_{a_{0}}\cap U_{a_{1}}$, the lifting $\tilde{\rho}_{a_{0}}(g)$ differs from $\tilde{\rho}_{a_{1}}(g)$ by a phase factor. We also have $\partial \omega =\delta\eta$, more explicitly:
\beq
\begin{split}
    \omega_{b_{1}}(g,h)\omega_{b_{2}}(g,h)^{-1}&=(\delta\eta)_{b_{1}b_{2}}(g,h)
    \\&=\eta_{d_{0}(b_{1}),d_{0}(b_{2})}(h)\eta_{d_{1}(b_{1}),d_{1}(b_{2})}(gh)^{-1}\eta_{d_{2}(b_{1}),d_{2}(b_{2})}(g)
\end{split}
\eeq
I.e. $\omega$ is only defined on each patch in $U_{2,0}$ and on their intersections (i.e. $U_{2,1}$) there is a transition function to connect them, which shifts $\omega$ by a coboundary $\delta\eta$. The data $[\omega,\eta]$ defines an element in $\rH^{2}_{\diff}(G;\U)$.

We note that if $\cH\simeq\bbC^{n}$, then for each $a\in I_{1}$ we define $\lambda_{a}(g):=(\det\tilde{\rho}_{a}(g))^{\frac{1}{n}}$. This is well-defined because $U_{\bullet,\bullet}$ is a good covering. Thus, one can shift $\tilde{\rho}_{a}(g)$ by $\lambda_{a}(g)$ such that $\det(\tilde{\rho}_{a}(g))=1$. This amounts to change $\omega_{b}$ by $(\delta\lambda)_{b}$ and it does not affect the resulting class in $\rH^{2}_{\diff}(G;\U)$. Therefore, by taking determinant on Eq.~\eqref{eq:local_2_cocycle} we have 
\beq
(\omega_{b})^{n}=1
\eeq
Similarly we have $(\eta_{a_{0},a_{1}})^{n}=1$. Thus, we can only realize the $n$-torsion subgroup of $\rH^{2}_{\diff}(G;\U)$ on $\bbC^{n}$.
\end{example}

To avoid complicated notations, we will henceforth suppress the subscripts in the differentiable group cohomology elements.

\medskip

\noindent
\textbf{Comparison of cohomology theories.}
\medskip

A priori, the different version of group cohomology, i.e. $\rH^{*}_{B}(G;\U),\rH^{*}_{loc,s}(G;\U)$ and $\rH^{*}_{\diff}(G;\U)$ may not coincide with each other.
Remarkably, for finite-dimensional Lie groups and suitable coefficients,
they are in fact isomorphic.

\begin{theorem}
[Corollary~IV.9 and Remark~IV.13 of Ref.~\cite{Wagemann2011cohomology}]
Let $G$ be a finite-dimensional Lie group acting smoothly on $\U$.
Then there are natural isomorphisms
\beq
\rH^{*}_{loc,s}(G;\U)
\;\simeq\;
\rH^{*}_{\mathrm{B}}(G;\U)
\;\simeq\;
\rH^{*}_{\diff}(G;\U) .
\eeq
\end{theorem}

Although these cohomology theories are isomorphic, the differentiable
group cohomology $\rH^{*}_{\diff}(G;\U)$ is often the most convenient
framework for constructing anomaly indices, see for instance
Ref.~\cite{kapustin2024anomalous}.
In most part of this paper, we therefore write
\[
\rH^{*}(G;\U) := \rH^{*}_{\diff}(G;\U) .
\]
for a Lie group $G$.

\medskip

We now collect several useful properties of differentiable group cohomology.
Proofs can be found in
Refs.~\cite{brylinski2000differentiable,Wagemann2011cohomology}.

\begin{proposition}\label{prop:discrete}
If either $G$ or $A$ is discrete, then
\beq
\rH^{*}_{\diff}(G;A) \simeq \rH^{*}(BG;A) ,
\eeq
where the right-hand side denotes the singular cohomology
of the classifying space $BG$.
\end{proposition}

\begin{proposition}
If $G$ is compact, then
\beq
\rH^{n}_{\diff}(G;\R) = 0 ,
\qquad n>0 .
\eeq
\end{proposition}

As a consequence of the Bockstein homomorphism, we obtain:

\begin{corollary}
Let $G$ be a compact Lie group. Then
\beq
\rH^{n}_{\diff}(G;\U)
\simeq
\rH^{n+1}(BG;\z),
\qquad n\ge 1 .
\eeq
\end{corollary}

\begin{proposition}
[K\"unneth formula, Appendix~B of Ref.~\cite{Cheng2016translation}]
Let $G$ and $H$ be finite-dimensional Lie groups (including discrete groups).
Then
\beq
\rH^{n}_{\diff}(G\times H;\U)
\simeq
\bigoplus_{p+q=n}
\rH^{p}_{\diff}\bigl(G;\rH^{q}_{\diff}(H;\U)\bigr) .
\eeq
\end{proposition}

\section{Proof to Theorem \ref{thm:main} for Lie groups}\label{sec:Proof_Lie}
Here we recall the main idea of constructing the anomaly index \cite{kapustin2024anomalous} for symmetry action $\alpha:G\to\G^{al}$, where $\G^{al}$ is a subgroup of $\G^{lp}$ with $O(r^{-\infty})$-tails. Similarly, one defines $\A^{al}$ to be a dense subalgebra of $\A$ with $O(r^{-\infty})$-tails and the unitary group in $\A^{al}$ is denoted by $\cU^{al}$.
It turns out $\A^{al}$ and $\G^{al}$ has nice smooth structure and thus it is a natural candidate for Lie group actions on spin chains.

The idea we outlined in Sec.~\ref{sec:sym_action} still applies: anomalies are obstructions to localize the symmetry to the right-half chain.  The only caveat is that if one can choose a smooth decomposition (i.e. $\alpha^{R}_{g}$ depends on $g$ smoothly). It is shown in Ref.~\cite{kapustin2024anomalous} that such decomposition is always possible on $U_{1,0}$ (see Lemma A.5 of \cite{kapustin2024anomalous}). Similarly, on $U_{2,0}$ we have
\beq
\alpha_{g}^{R}\alpha_{h}^{R}=\Ad_{V_{g,h}}\alpha^{R}_{gh},\quad (g,h)\in U_{2}^{(b)}\in U_{2,0}
\eeq
Furthermore on $U_{3,0}$ we have
\beq
\omega_{c}(g,h,k):=V_{g,h}V_{gh,k}V_{g,hk}^{-1}\alpha_{g}^{R}(V_{h,k})^{-1}, \quad (g,h,k)\in U^{(c)}_{3}\in U_{3,0}
\eeq

We recall that for any contractible space $W\subseteq\R^{D}$ (where $D\geqslant1$), the family of unitaries $W\to\cU^{al}$ is called smooth if $\psi\circ u_{w}$ is a smooth function on $W$ for any state $\psi$. Thus, it is easy to show that the (log-)determinant function $\Delta_{\tau}$ (Eq.~\eqref{eq:det_log}) defines a smooth map for a smooth family $W\to\cU^{al}$ since $\tau(u^{-1}\rd u)$ is a smooth $\bbC$-valued 1-form on $W$. We will take $W$ to be components of $U_{\bullet,\bullet}$ below.

On $U_{2,0}$, we can always make a gauge fixing such that $\Delta_{\tau}(V_{g,h})=0\,\quad\mathrm{mod}\,\,\z[n^{-1}]$ by redefining $V'_{g,h}:=\lambda_{g,h}^{-1}V_{g,h}$ where $\lambda_{g,h}:=\det_{\tau}(V_{g,h})$, which is smooth on $U_{2,0}$ as well. With this choice, we have $\Delta_{\tau}(\omega)=0,\,\,\mathrm{mod}\,\,\z[n^{-1}]/\z$. Similar gauge fixing is possible for other data in the differentiable group cohomology. Thus the classification reduces to
\beq
\rH^{3}_{\diff}(G;\z[n^{-1}]/\z)\simeq\rH^{3}(G;\z[n^{-1}]/\z)
\eeq
where the right-hand-side the singular cohomology of $BG$.

\bibliography{lib}
\end{document}